\documentclass[showpacs,keywords,preprintnumbers,amsmath,amssymb]{revtex4}
 \usepackage{bm}
\usepackage[dvips]{graphicx}
\usepackage{latexsym,amsmath,amssymb}
\usepackage{epsfig,subfigure}
\usepackage{color}

\begin{document}

\title[]{Particle Acceleration in Kerr-(anti-) de Sitter Black Hole Backgrounds}


\author{Yang Li}
\thanks{E-mail: liyang09@lzu.cn}

\author{Jie Yang}
\thanks{E-mail: yangjiev@lzu.edu.cn, Corresponding author}

\author{Yun-Liang Li}
\thanks{E-mail: liyl09@lzu.cn}

\author{Shao-Wen Wei}
\thanks{E-mail: weishw@lzu.edu.cn}

\author{Yu-Xiao Liu}
\thanks{E-mail: liuyx@lzu.edu.cn}
 \affiliation{Institute of Theoretical Physics, Lanzhou University,
           Lanzhou 730000, China}


\begin{abstract}
{Ban\~{a}dos, Silk and West (BSW) proved that Kerr black holes could act
as particle accelerators with arbitrarily high center-of-mass energy, if the two conditions are satisfied: (1) These black holes are extremal; (2) one of the colliding particles has critical angular momentum.
In this paper, we extend the research to the cases of Kerr-(anti-) de Sitter black holes and find that the cosmological constant has an important effect on the result. In order for the case of Kerr-anti-de Sitter black holes (with negative cosmological constant) to get arbitrary high center-of-mass energy, we need an additional condition besides the above two for Kerr ones.
While, for the case of general Kerr-de Sitter black holes (with positive cosmological constant), the collision of two particles can take place on the outer horizon of the black holes and the center-of-mass energy of collision can blow up arbitrarily if the above second condition is satisfied. Hence, non-extremal Kerr-de Sitter black holes could also act
as particle accelerators with arbitrarily high center-of-mass energy.}
\end{abstract}


\maketitle


\section{Introduction}

{\color{blue} Two years ago}, Ban\~{a}dos, Silk and West reported a process (BSW process) that two particles may collide on the horizon of the extremal Kerr black hole with the arbitrarily high center-of-mass (CM) energy \cite{Banados1}. { Although it has been pointed out in Refs. \cite{Berti,Jacobson} that} there are astrophysical limitations preventing a Kerr black hole to be extremal, and the gravitational radiation and backreaction effects should be counted in this process, similar processes { have been found} in other kinds of black holes or naked singularities and the BSW process of the Kerr black hole { has been studied} more deeply \cite{Jacobson, Grib1, Lake, Grib2, Wei1, Grib3, Zaslavskii1, Wei2, Mao, Harada, Grib4, Banados2, Patil}. On the other hand, { a general analysis of this BSW process has been done for rotating black holes \cite{Zaslavskii2} and for most general black holes \cite{Zaslavskii3,WuPRD}}.
Some efforts have also been made to draw some implications concerning the effects of gravity generated by colliding particles in \cite{Kimura}.

In this paper, we investigate the BSW process of the Kerr-(anti-) de Sitter black hole, and our goal is to see { the effect of the cosmological constant on} the BSW process. There are good reasons to believe that { our results  can be reduced to the ones of BSW given in \cite{Banados1} as the cosmological constant turns to zero.} Besides, because the Kerr-(anti-) de Sitter black hole does not have a simple horizon structure as the previous studied black holes, we have to { use a different method to study} the BSW process.

This paper is organized as follows. In Sec. \ref{SecBHs},
we { study} the horizon structure of Kerr-(anti-) de Sitter black holes.
In Sec. \ref{SecCMEnergy}, we { calculate}
the CM energy of the particle collision on the horizon of the black holes,
and derive the critical angular momentum to make the CM
energy to blow up. In Sec. \ref{SecRadialMotion},
we find the BSW process requirements for the black hole
and the colliding particle from the geodesic motion of the colliding particle.
The conclusion is given in the last section.

\section{Extremal Kerr-(anti-) de Sitter black holes}
\label{SecBHs}

In this section, we would like to study the extremal the Kerr-(anti-) de Sitter black holes. First, the vacuum metric of the Kerr-anti-de Sitter (Kerr-AdS) black holes in Boyer-Lindquist coordinate system {with units $c=G=1$} is given by
\begin{eqnarray}
ds^2 = -\frac{\Delta_\text{r}}{\rho^2}(dt - \frac{a}{\Sigma}\text{sin}^2\theta d\phi)^2 + \frac{\rho^2dr^2}{\Delta_r} + \frac{\rho^2d\theta^2}{\Delta_\theta} + \frac{\Delta_\theta\text{sin}^2\theta}{\rho^2}(adt - \frac{r^2 + a^2}{\Sigma}d\phi)^2,\label{metric}
\end{eqnarray}
where
\begin{eqnarray}
\Delta_r &=& (r^2 + a^2)(1 + \frac{r^2}{l^2}) - 2Mr,\\
\Delta_\theta &=& 1 - \frac{a^2\text{cos}^2\theta}{l^2},\\
\rho^2 &=& r^2 + a^2\text{cos}^2\theta,\\
\Sigma &=& 1 - \frac{a^2}{l^2},
\end{eqnarray}
$M$ is related to the mass of the black hole, $a$ is related to the black hole's spin angular momentum per mass by $a = J/M$, and $l^2$ is related to the cosmological constant $\Lambda$ by $l^{-2} = -\Lambda/3$. And for the Kerr-de Sitter (Kerr-dS) black hole, the form of the vacuum metric will remain the same, but $l^2$ in $\Delta_r$ and $\Sigma$ should be replaced by $-l^2$.

The horizon $r_\text{h}$ is given by $\Delta_r|_{r=r_\text{h}} = 0$. We can make a comparison of coefficients in
$\Delta_r = l^{-2}r^4 + {(1 + a^2l^{-2}) r^2} - 2Mr + a^2$
$= l^{-2}\prod^4_{i=1}(r-r_{\text{h}i})$ for the Kerr-AdS case, where $r_{\text{h}i}$ ($i=1, 2, 3, 4$) denotes the zeros of $\Delta_r$ \cite{Hackmann}. From this comparison we can conclude that for the Kerr-AdS black hole, there are two separated positive horizons at most, { and $\Delta_r$ is positive} outside the outer horizon of the Kerr-AdS black hole. In the same way, we can conclude for the Kerr-dS black hole { that} there are three separated positive horizons at most, and $\Delta_r$ is negative outside the outer horizon of the black hole. For the Kerr-AdS (Kerr-dS) black hole, when two horizons of the black hole coincide, the black hole is extremal.

If consider the extremal Kerr-AdS black hole, we have to make a comparison of coefficients in $\Delta_r = l^{-2}r^4 + {(1 + a^2l^{-2}) r^2} - 2Mr + a^2 = (r-x)^2\left(a_2r^2+a_1r+a_0\right)$ with $a_0, a_1, a_2$ being real \cite{Hackmann}. From this comparison we can get
\begin{eqnarray}
\frac{a^2 l^2}{x^2} - 3 x^2 &=& a^2  + l^2,\label{Mads1}\\
 2x^3-2 \frac{a^2 l^2}{x} &=& -2 Ml^2 \label{Mads2}.
\end{eqnarray}
In these equations, $x$ is positive and related to the coincided horizon of the extremal Kerr-AdS black hole. Then $x$ can be solved as
\begin{eqnarray}
x = \sqrt{-\frac{a^2}{6}-\frac{l^2}{6}+\frac{1}{6} \sqrt{a^4+14 a^2 l^2+l^4}}. \label{xads}
\end{eqnarray}
Analogously we can get these equations for the Kerr-dS case:
\begin{eqnarray}
\frac{a^2 l^2}{x^2} +3 x^2 = -a^2  + l^2, \label{Mds1}\\
{2x^3+2 \frac{a^2 l^2}{x} = 2 Ml^2}\label{Mds2},
\end{eqnarray}
where $x$ is positive and related to { the} coincided horizon of the extremal Kerr-dS black hole. Also $x$ can be solved as
\begin{eqnarray}
x_1 &=& \sqrt{-\frac{a^2}{6}+\frac{l^2}{6}-\frac{1}{6} \sqrt{a^4-14 a^2 l^2+l^4}}{,} \label{xds1}\\
x_2 &=& \sqrt{-\frac{a^2}{6}+\frac{l^2}{6}+\frac{1}{6} \sqrt{a^4-14 a^2 l^2+l^4}}.\label{xds2}
\end{eqnarray}
These two solutions are both { the} coincided horizons of the extremal Kerr-dS black hole. By computing {$d^2\Delta_r/dr^2$} at $r=x_1$ and $r=x_2$, we can see that { $x_1$ and $x_2$ are related to the inner and outer coincided horizons, respectively.}

We can also solve $M$ and $a$ from Eqs. (\ref{Mds1}) and (\ref{Mds2}):
\begin{eqnarray}
M &=& \frac{x(l^2-x^2)^2}{l^2 (l^2 + x^2)},\\
a^2 &=& \frac{x^2 \left(l^2-3 x^2\right)}{l^2+x^2},
\end{eqnarray}
from which we can see { that} $x^2 {<} l^2/3$ and $a^2$ has a range of $0{<} a^2{<} \left(7-4 \sqrt{3}\right) l^2$. So for an extremal Kerr-dS black hole, there are { upper limits for the extremal horizon and the angular momentum of the black hole}. While for a Kerr-AdS black hole, $a$ also { has an upper limit $l$}. When $a$ reaches $l$, the metric will be singular, and when $a$ exceeds $l$, the Kerr-AdS black hole { will be unstable} due to the superradiance \cite{Hawking}.

\section{The CM Energy of the collision on the horizon of the Kerr-(anti-) de Sitter black hole }
\label{SecCMEnergy}

To investigate the CM energy of the collision on the horizon of the Kerr-(anti-) de Sitter Black Hole, we have to { derive} the 4-velocity of the colliding particle. And we only study the particle motion on the equatorial plane ($\theta = \frac{\pi}{2}$, $\rho^2 = r^2$).\\
\indent The generalized momenta $P_\mu$ can be given as
\begin{equation}
P_\mu=g_{\mu \nu}\dot{x}^\nu,
\end{equation}
where the dot denotes { the} derivative with respect to { the} affine parameter $\lambda$ and $\mu, \nu = t, r, \phi, \theta$. Thus, in equatorial motion, generalized momenta $P_t$ and $P_\phi$ are turned out to be
\begin{eqnarray}
P_t &=& g_{tt} \dot{t} + g_{t \phi} \dot{\phi},\label{4-veloE} \\
P_\phi &=& g_{\phi \phi} \dot{\phi} + g_{t \phi} \dot{t}.\label{4-veloL}
\end{eqnarray}
$P_t$ and $P_\phi$ are constants of motion. In fact, { $P_t$ and $P_\phi$ correspond to the test particle's energy per unit mass $E$ and the angular momentum parallel to the symmetry axis per unit mass $L$, respectively.} And in the following discussion we will just regard these two constants of motion as $E\equiv P_t$ and $L\equiv P_\phi$ \cite{Hackmann}.

The affine parameter $\lambda$ can be related to the proper time by $\tau = \mu \lambda$, where $\tau$ is given by the normalization condition $-{\mu}^2=g_{\mu \nu}\dot{x}^\mu \dot{x}^\nu$ with $\mu = 1$ for timelike geodesics and $\mu=0$ for { null geodesics}. For a timelike geodesic, the affine parameter can be identified with the proper time, and thus from Eqs. (\ref{4-veloE}) and (\ref{4-veloL}), we can solve 4-velocity components $\dot{t}$ and $\dot{\phi}$:
\begin{eqnarray}
\frac{dt}{d\tau}&=&\frac{E\left(\left(a^2 + r^2\right)^2-\Delta_r a^2\right)+L \Sigma  a\left(\Delta_r-a^2-r^2\right)}{r^2 \Delta_r}, \label{tt} \\
\frac{d\phi}{d\tau}&=&\frac{E\Sigma  a\left(r^2+a^2-\Delta_r\right)+L \Sigma ^2\left(\Delta_r -a^2 \right)}{r^2 \Delta_r}.\label{phit}
\end{eqnarray}
For { the remained component} $\dot{r} = \frac{dr}{d\tau}$ of the equatorial motion, we can obtain it from the Hamilton-Jacobi equation of the { timelike} geodesic
\begin{eqnarray}
\frac{\partial S}{\partial \tau} = -\frac{1}{2}g^{\mu \nu}\frac{\partial S}{\partial x^\mu}\frac{\partial S}{\partial x^\nu}\label{HYE}
\end{eqnarray}
with the ansatz
\begin{eqnarray}
S = \frac{1}{2}\tau - Et + L\phi + S_r(r),
\end{eqnarray}
where $S_r(r)$ is the function of $r$. Inserting the ansatz into (\ref{HYE}), and with the help of the metric (\ref{metric}), we get
\begin{eqnarray}
\left(\frac{dS_r(r)}{dr} \right)^2 &=& \frac{K^2-\Delta_r\left(r^2+( L \Sigma -a E)^2\right)}{\Delta_r^2},\\
K &=& a^2 E+E r^2-a L \Sigma.
\end{eqnarray}
On the other hand, we have
\begin{eqnarray}
\frac{dS_r(r)}{dr} = P_r= g_{rr}\dot{r} = \frac{r^2}{\Delta_r}\dot{r}.
\end{eqnarray}
Thus, we get the square of the radial 4-velocity component
\begin{eqnarray}
\left(\frac{dr}{d\tau}\right)^2 &=& \frac{K^2-\Delta_r\left(r^2+( L \Sigma -a E)^2\right)}{r^4}.\label{rt}
\end{eqnarray}
Here we have obtained all nonzero 4-velocity components for the equatorial motion geodesic. Next we would like to study the CM energy of the two-particle collision in the backgrounds of Kerr-(anti-) de Sitter black holes. We assume that the two particles have the angular mentum per unit mass $L_1, L_2$ and energy per unit mass $E_1, E_2$, respectively. For simplicity, the particles in consideration have the same rest mass $m_0$. { The expression of the CM energy $E_{\text{CM}}$ of this two-particle collision is given by \cite{Banados1}}
\begin{eqnarray}
E_{\text{CM}} = \sqrt{2} m_0 \sqrt{1-g_{\mu \nu}u^{\mu}_1u^{\nu}_2},
\end{eqnarray}
where $u^{\mu}_1, u^{\nu}_2$ are the 4-velocity vectors of the two particles ($u=(\dot{t},\, \dot{r},\, 0,\, \dot{\phi})$). With the help of { Eqs. (\ref{tt}, \ref{phit}, \ref{rt})}, we obtain the CM energy
\begin{eqnarray}
\frac{1}{2m_0^2}E_{\text{CM}}^2 &=& \frac{1}{r^2 \Delta_r}\left((E_2 L_1+E_1 L_2) \Sigma  a\left(\Delta_r-a^2-r^2\right)+L_1 L_2 \Sigma ^2\left(a^2-\Delta_r\right)\right)+ \nonumber \\
&&\frac{1}{r^2 \Delta_r}\left(E_1 E_2\left(\left(a^2 + r^2\right)^2-a^2 \Delta_r \right)+r^2 \Delta_r-H_1H_2\right),\label{Ecm}
\end{eqnarray}
where
\begin{eqnarray}
H_i &=& \sqrt{E_i^2\left(a^2 + r^2\right)^2-a^2 E_i^2 \Delta_r-r^2 \Delta_r-2 a E_i L_i \left(a^2+r^2-\Delta_r\right) \Sigma +L_i^2 \left(a^2-\Delta_r\right) \Sigma ^2} \nonumber \\
&&(i=1, 2).
\end{eqnarray}
For simplicity, we can rescale the CM energy as ${ {\bar{E}_{\text{CM}}^2}} \equiv \frac{1}{2m_0^2}E_{\text{CM}}^2$. We would like to see $\bar{E}_{\text{CM}}^2$ when the particles collide { on the horizon}. So we have to make $\Delta_r=0$ at Eq. (\ref{Ecm}). The denominator of $\bar{E}_{\text{CM}}^2$ is zero, and the numerator of it is
\begin{eqnarray}
K_1K_2-\sqrt{K_1^2}\sqrt{K_2^2},\\
K_i=K|_{E=E_i, L=L_i},\,\,  i=1, 2.
\end{eqnarray}
When $K_1K_2\geq0$, the numerator will be zero and the value of $\bar{E}_{\text{CM}}^2$ on the horizon will be undetermined; but when $K_1K_2<0$, the numerator will be negative finite value and $\bar{E}_{\text{CM}}^2$ on the horizon will be negative infinity. { So it should have $K_1K_2\geq0$}, and for the CM energy on the horizon, we have to compute the limiting value of equation (\ref{Ecm}) as $r \rightarrow r_\text{h}$, where $r_\text{h}$ is related to the horizon of the black hole.

We can make $\Delta_r=(r-r_\text{h})\left(a_3r^3+a_2r^2+a_1r+a_0\right)$ in Eq. (\ref{Ecm}). Then we expand Eq. (\ref{Ecm}) at $r_\text{h}$, which is the horizon under the consideration. When $r=r_\text{h}$, the remaining term in the expansion of (\ref{Ecm}) is the zero-order term. In fact, the zero-order term is of the lowest order in the expansion of (\ref{Ecm}). So the limiting value of $\bar{E}_{\text{CM}}^2$ as $r \rightarrow r_\text{h}$ is given by the zero-order term of (\ref{Ecm}) as
\begin{eqnarray}
\bar{E}_{\text{CM}}^2(r \rightarrow r_\text{h})=\frac{A}{2K_1K_2},
\end{eqnarray}
where
\begin{eqnarray}
A &=& \Big[\left(a^2+r_\text{h}^2 \right)(E_1+E_2) -a \Sigma (L_1+L_2) \Big]^2 +r_\text{h}^2 \Sigma ^2(E_2 L_1-E_1 L_2)^2.
\end{eqnarray}
When $K_1=a^2 E_1+E_1 r_\text{h}^2-a L_1 \Sigma=0$, $A$ will be
\begin{eqnarray}
A=\frac{\left(a^2+E_1^2 r_\text{h}^2\right) K_2^2}{a^2}.
\end{eqnarray}
So we can see that when $K_1=0$ and $K_2\neq0$, the CM energy on the horizon will blow up. We will call the angular momentum per unit mass that make $K_i=0$ the critical angular momentum $L_{\text{C}i}$, and $L_{\text{C}i}$ is given as
\begin{eqnarray}
L_{\text{C}i}=\frac{(a^2 + r_\text{h}^2)E_i}{a\Sigma }\label{Lc}, \,\, i=1,2.
\end{eqnarray}
\indent {We can also prove that when $K_1=0$ and $K_2=0$, the CM energy will not blow up. So if we need the CM energy to be arbitrarily high, one of the colliding particles must have the critical angular momentum and the other particle must not have the critical angular momentum.}

We can see that the critical angular momentum depends on the horizon $r_\text{h}$, and when we consider different horizons of the black hole, the critical angular momentums { correspond to} the horizons will be different. This result can reduce to the critical angular momentum { for the case of} the Kerr black hole when the cosmological constant { becomes} zero.

{ In order to get arbitrarily high CM energy on the horizon of the Kerr-AdS(dS) black hole,} the colliding particle with the critical angular momentum must be able to reach the outer horizon of the black hole. We will study this part in next section.

\section{The radial motion of the particle with the critical angular momentum near the outer horizon of the black hole.}
\label{SecRadialMotion}

In this section, we will study { the conditions under which} the particle with the critical angular momentum can reach the outer horizon of the black hole. { In order for a particle to reach the horizon of the black hole, the square of the radial component of the 4-velocity $\left(\frac{dr}{d\tau}\right)^2$ in Eq. (\ref{rt}) } has to be positive { in the neighborhood outside of} the black hole's horizon.

We denote $\left(\frac{dr}{d\tau}\right)^2$ as $R(r)$. Obviously, when the particle has the critical angular momentum, $R(r)=0$ on the horizon of the Kerr-AdS or Kerr-dS black hole. So if the particle with the critical angular momentum can reach the horizon of the black hole, the derivative of $R(r)$ { with respect to} $r$ must be positive at the horizon $r_\text{h}$, i.e.,
\begin{eqnarray}
\frac{d R(r)}{dr}|_{r=r_\text{h}}>0.
\end{eqnarray}
{ Before doing the computation}, we would like to make a parameter replacement
\begin{eqnarray}
M=\frac{((r_\text{h}')^2 + a^2)(1 + \frac{(r_\text{h}')^2}{l^2})}{2 (r_\text{h}')} \label{Mrhads}
\end{eqnarray}
for the Kerr-AdS black hole, and
\begin{eqnarray}
M=\frac{((r_\text{h}')^2 + a^2)(1 - \frac{(r_\text{h}')^2}{l^2})}{2 (r_\text{h}')} \label{Mrhds}
\end{eqnarray}
for the Kerr-dS black hole. After this parameter replacement, $r_\text{h}'$ will be the horizon of the black hole ($\Delta_r|_{r=r_\text{h}'}=0$). So we can start to discuss the black hole's horizon $r_\text{h}'$ and let $r_\text{h}'$ be identified with $r_\text{h}$. Thus we will only write $r_\text{h}$ in the following discussion. For the Kerr-dS black hole, $r_\text{h}$ must not exceed $l$ to avoid a negative $M$. After this parameter replacement, computing $\frac{d R(r)}{dr}$ at $r=r_\text{h}$ { for the particle} with the critical angular momentum will give
\begin{eqnarray}
\frac{d R(r)}{dr}|_{r=r_\text{h}}=W_{\text{AdS}}\cdot B
\end{eqnarray}
for the Kerr-AdS black hole, and
\begin{eqnarray}
\frac{d R(r)}{dr}|_{r=r_\text{h}}=W_{\text{dS}}\cdot B \label{dRds}
\end{eqnarray}
for the Kerr-dS black hole, where
\begin{eqnarray}
W_{\text{AdS}} &=& a^2 \left({l^2-r_\text{h}^2}\right){-}r_\text{h}^2 \left(l^2+3 r_\text{h}^2\right),\\
W_{\text{dS}} &=& a^2 \left(l^2+r_\text{h}^2\right)-r_\text{h}^2 \left(l^2-3 r_\text{h}^2\right),\\
B &=& \frac{ a^2+E^2 r_\text{h}^2 }{a^2 r_\text{h}^3l^2}.
\end{eqnarray}
Notice that because $B>0$, whether $\frac{d R(r)}{dr}|_{r=r_\text{h}}$ is positive only depends on the sign of $W_{\text{AdS}}$ or $W_{\text{dS}}$. Both $W_{\text{dS}}$ and $W_{\text{AdS}}$ only depend on the parameters of the black hole. Next, { we will discuss the cases of the Kerr-AdS and Kerr-dS black holes respectively.}

\subsection{The Kerr-AdS case}

\begin{figure}
\center
\includegraphics[width=120mm]{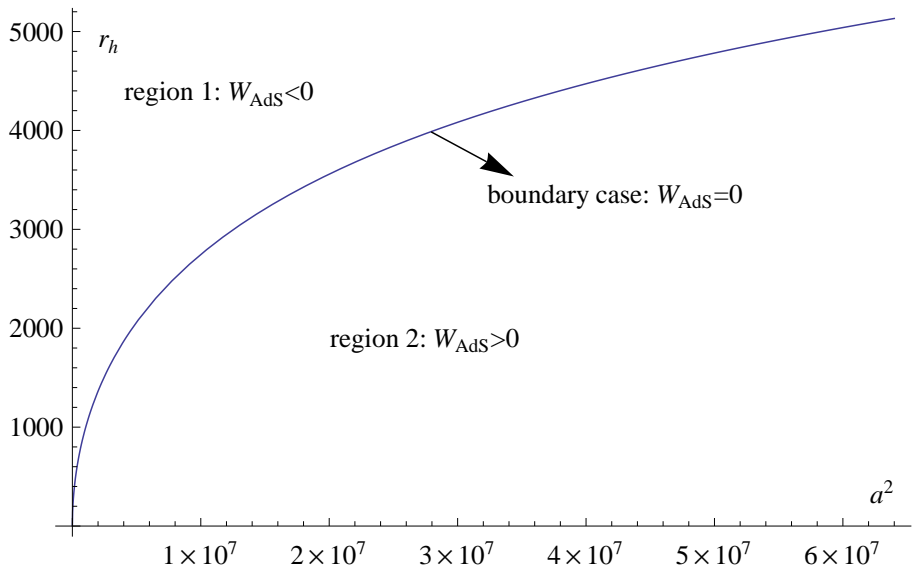}
\caption{{ The shape of $r_\text{h}(a^2,l^2)$ in (\ref{rhads}) with $l=10^4$.} The point $(a^2,r_\text{h})$ in region 2 will make $\frac{d R(r)}{dr}|_{r=r_\text{h}}$ positive, which means the particle with critical angular momentum can reach the horizon $r_\text{h}$.}
\label{Wregionads}
\end{figure}

In Kerr-AdS black hole case, { by solving $W_{\text{AdS}}=0$, we get}
\begin{eqnarray}
r_\text{h}= \sqrt{-\frac{a^2}{6}-\frac{l^2}{6}+\frac{1}{6} \sqrt{a^4+14 a^2 l^2+l^4}}. \label{rhads}
\end{eqnarray}
{ We draw the shape of $r_\text{h}(a^2,l^2)$ in (\ref{rhads}) with $l=10^4$ in Fig. \ref{Wregionads}, in which every point is related to a combination of the black hole horizon $r_\text{h}$ and the black hole spin $a$, and the point $(a^2,r_\text{h})$ on the line means that the corresponding $W_{\text{AdS}}$ is zero.} Because $W_{\text{AdS}}$ is a continuous function of $r_\text{h}$ and $a_2$, the different regions in Fig. \ref{Wregionads} separated by the line ($W_{\text{AdS}}=0$) relate to different signs of $W_{\text{AdS}}$. So we call $W_{\text{AdS}}=0$ is the boundary case. We can verify that above the line, $r_\text{h}$ and $a^2$ make $W_{\text{AdS}}<0$; and below the line, $r_\text{h}$ and $a^2$ make $W_{\text{AdS}}>0$.

When $W_{\text{AdS}}>0$, the particle with critical angular momentum can reach the horizon $r_\text{h}$. But we have to make sure that $r_\text{h}$ is the outer horizon of the black hole. Notice that Eq. (\ref{rhads}) is just the same as Eq. (\ref{xads}), which means $r_\text{h}$ in the boundary case of $\frac{d R(r)}{dr}|_{r=r_\text{h}}=0$ is the extremal horizon $x$ of the Kerr-AdS black hole. This means { when the point $(a^2,~r_\text{h})$ is on the line} in Fig. \ref{Wregionads}, the black hole is extremal. But when $(a^2,~r_\text{h})$ is off the line, is the black hole extremal or not? To answer that, we must make the parameter replacement (\ref{Mrhads}) in the extremal horizon equations (\ref{Mads1}, \ref{Mads2}), and solve { them} for $r_\text{h}$. Eq. (\ref{rhads}) must be one of the solutions, and { this solution is related to the extremal horizon}. Recall that the Kerr-AdS black hole can have two positive horizons at most. So if the black hole is extremal, these two horizons must coincide and the black hole will only have one horizon, namely $r_\text{h}$. Thus Eq. (\ref{rhads}) is the only solution. This means that only if $(a^2,r_\text{h})$ is on the line in Fig. \ref{Wregionads}, the Kerr-AdS black hole can be extremal.

So the line in Fig. \ref{Wregionads} can also be regarded as the boundary case of the horizon situation of the black hole. This is because when we pick { a point $(a^2,r_\text{h})$} off the line in Fig. \ref{Wregionads}, if we find $r_\text{h}$ is the inside horizon of the black hole, $r_\text{h}$ cannot turn into the outer horizon by crossing the other horizon or the number of the black hole horizons cannot change unless the point cross the line. In this case, when { the point $(a^2,r_\text{h})$ is above the line}, $r_\text{h}$ is the outer horizon and the black hole has two horizons; when $(a^2,r_\text{h})$ is on the line, $r_\text{h}$ is the extremal horizon and the black hole has only one extremal horizon; when $(a^2,r_\text{h})$ is below the line, $r_\text{h}$ is the inner horizon and the black hole has two horizons. This means that if we want the particle with critical angular momentum to reach the outer horizon of the Kerr-AdS black hole, the only chance is the black hole is extremal. Thus, we must choose { these points $(a^2,r_\text{h})$ on the line in Fig. \ref{Wregionads}}. But in this boundary case, $\frac{d R(r)}{dr}|_{r=r_\text{h}}=0$ and we must calculate $\frac{d^2 R(r)}{dr^2}|_{r=r_\text{h}}$:
\begin{eqnarray}
\frac{d^2 R(r)}{dr^2}|_{r=r_\text{h}}=G \cdot E^2 + J,\label{d2R2}
\end{eqnarray}
where
\begin{eqnarray}
G &=&\frac{a^8+a^6 l^2-96 a^4 l^4-65 a^2 l^6-5 l^8 }{54 a^2} \nonumber\\
&+&\frac{\sqrt{a^4+14 a^2 l^2+l^4} \left(-a^6+6 a^4 l^2+30 a^2 l^4+5 l^6\right)}{54 a^2},
\end{eqnarray}
and
\begin{eqnarray}
J &=& \frac{1}{18}\left(-a^6-a^4 l^2+13 a^2 l^4+l^6\right) \nonumber\\
&+&\frac{1}{18}\left(a^4-6 a^2 l^2-l^4\right) \sqrt{a^4+14 a^2 l^2+l^4}.
\end{eqnarray}
In above calculation we { have already used} Eq. (\ref{rhads}). If $\frac{d^2 R(r)}{dr^2}|_{r=r_\text{h}}>0$, the particle with the critical angular momentum can reach the only horizon of the extremal Kerr-AdS black hole. It can be proved that if $\frac{d^2 R(r)}{dr^2}|_{r=r_\text{h}}>0$,
\begin{eqnarray}
&&0<a^2< \left(5 -2 \sqrt{5}\right)l^2,\label{addionads1}\\
&&E>\sqrt{\frac{-J}{G}}\label{addionads2}.
\end{eqnarray}
{And if $\frac{d^2 R(r)}{dr^2}|_{r=r_\text{h}}=0$, we can prove that $\frac{d^3 R(r)}{dr^3}|_{r=r_\text{h}}$ must be negative.}
Notice that the upper limit of black hole spin $a$ in Eq. (\ref{addionads1}) is still below $l$.

{ Now we summarize the result for the case of Kerr-AdS black hole and give a comparison to the case of Kerr black hole. We find that, for a non-extremal Kerr-AdS black hole, the particle with the critical angular momentum cannot reach the outer horizon of the black hole, which is the same with the case of Kerr black hole. However, for an extremal Kerr-AdS black hole, if the additional conditions (\ref{addionads1}) and (\ref{addionads2}) are satisfied, the particle with the critical angular momentum can reach the outer horizon of the black hole. While, for an extremal Kerr black hole, this process can always happen.}

\subsection{The Kerr-dS case}

\begin{figure}
\center
\includegraphics[width=120mm]{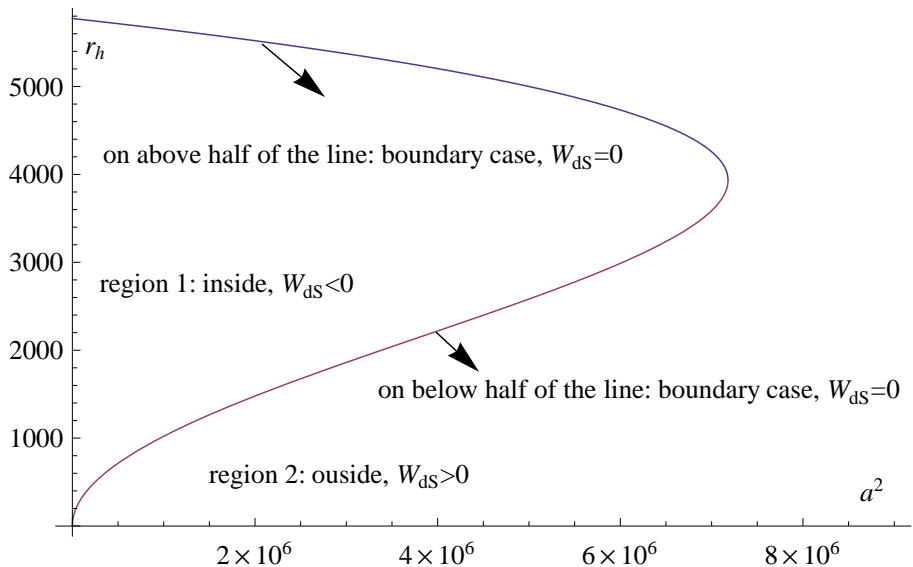}
\caption{{ The shapes of $r_{\text{h}1}(a^2,l^2)$ in (\ref{rhds1}) (the below line) and $r_{\text{h}2}(a^2,l^2)$ in (\ref{rhds2}) (the above line) with $l=10^4$.} In region 2, $(a^2,r_\text{h})$ can make $\frac{d R(r)}{dr}|_{r=r_\text{h}}>0$, which means that when $(a^2,r_\text{h})$ is in region 2, the particle with the critical angular momentum can reach the black hole horizon $r_\text{h}$. In the contrast, when $(a^2,r_\text{h})$ is in region 1, the particle with the critical angular momentum cannot reach the black hole horizon $r_\text{h}$.}
\label{Wregionds}
\end{figure}

Analogously to the Kerr-AdS case, we solve the boundary case $W_{\text{dS}}=0$ and get
\begin{eqnarray}
r_{\text{h}1} &=& \sqrt{-\frac{a^2}{6}+\frac{l^2}{6}-\frac{1}{6} \sqrt{a^4-14 a^2 l^2+l^4}}, \label{rhds1}\\
r_{\text{h}2} &=& \sqrt{-\frac{a^2}{6}+\frac{l^2}{6}+\frac{1}{6} \sqrt{a^4-14 a^2 l^2+l^4}}\label{rhds2}.
\end{eqnarray}
We draw these two boundary lines in Fig. \ref{Wregionds} with $l=10^4$. We can see that these two boundary lines join together at $a^2=\left(7-4 \sqrt{3}\right) l^2$. So actually there is only one boundary line in Fig. \ref{Wregionds}. { Like the Kerr-AdS case}, we verify that inside the boundary line, $(a^2,r_\text{h})$ makes $W_{\text{dS}}<0$; and outside the boundary line, $(a^2,r_\text{h})$ makes $W_{\text{dS}}>0$.

\begin{figure}
\center
\includegraphics[width=120mm]{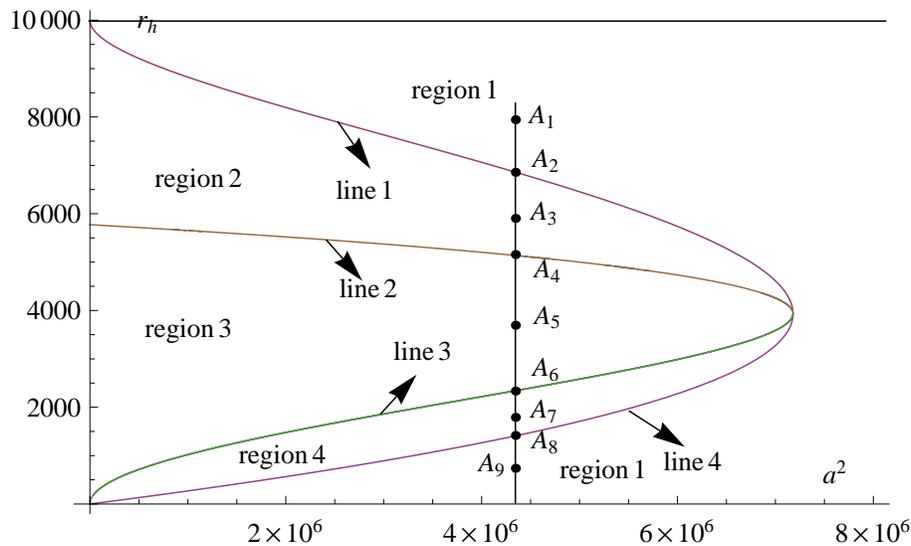}
\caption{Four boundary lines and {the two axes} separate the plane into four different regions. The horizontal line on the top in the figure refers to the upper limit of $r_\text{h}$.}
\label{extremerhds}
\end{figure}

When $W_{\text{dS}}>0$, we still have to make sure that $r_\text{h}$ is the outer horizon of the black hole. Notice that Eqs. (\ref{rhds1}) and (\ref{rhds2}) are the same as Eqs. (\ref{xds1}) and (\ref{xds2}). This means that the boundary line in Fig. \ref{Wregionds} is also the boundary line of the horizon situation of the black hole. But unlike the Kerr-AdS case, the Kerr-dS black hole can have three positive horizons at most and this means there are other boundary lines of the horizon situation of the black hole. To find them, we make the parameter replacement (\ref{Mrhds}) in Eqs. (\ref{Mds1}) and (\ref{Mds2}), and solve it for $r_\text{h}$. Obviously, Eqs. (\ref{rhds1}) and (\ref{rhds2}) are two solutions. And there are two other solutions which relate to the situations that the black hole is extremal but $r_\text{h}$ is not the extremal horizon. We draw all this boundary lines of the horizon situation of the black hole in Fig. \ref{extremerhds} with $l=10^4$.

\begin{figure}
\begin{center}
\includegraphics[width=170mm]{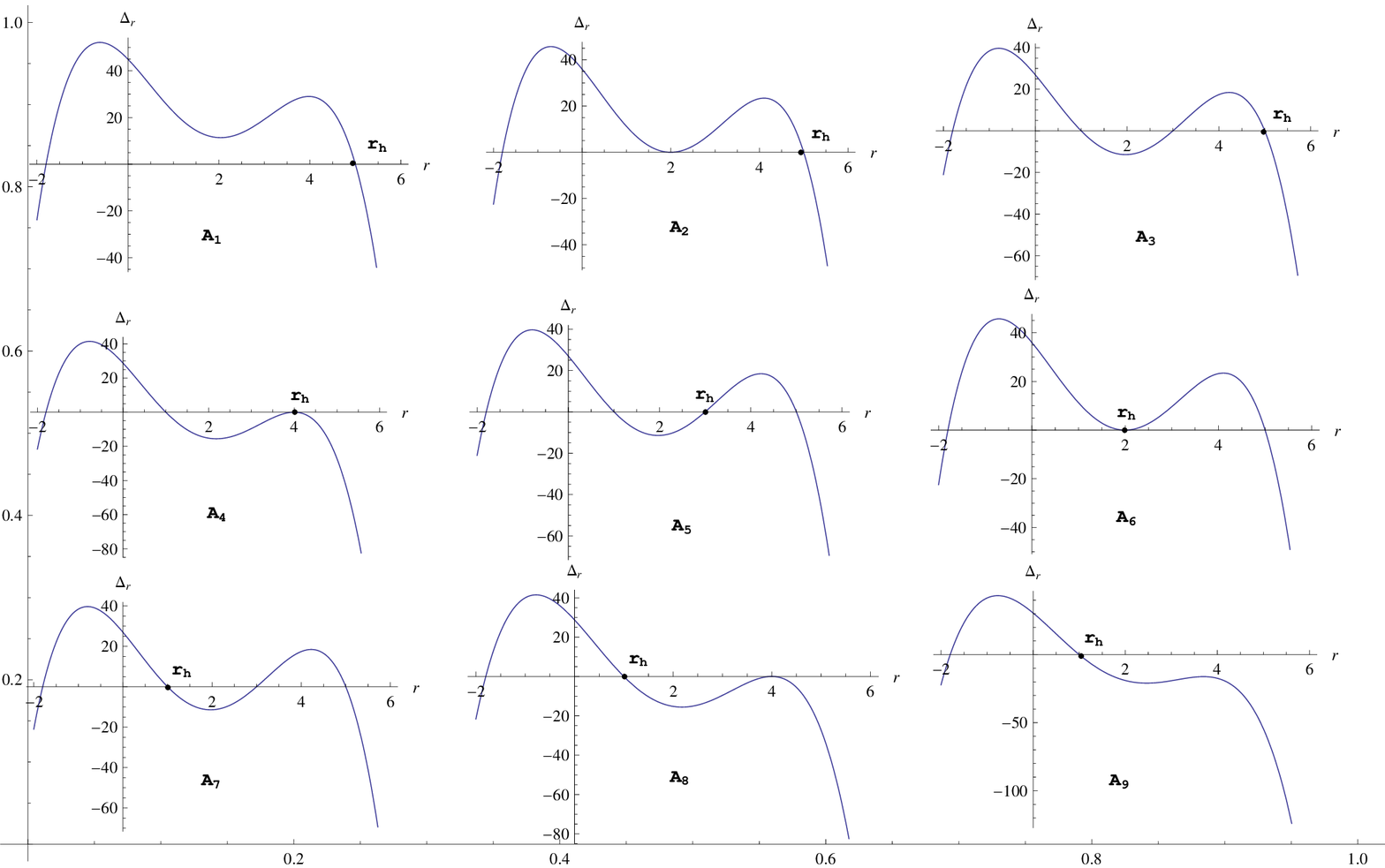}
\caption{{ The different types of the horizon situation of the Kerr-dS black hole.} The coordinates in this figure have no meanings. ${A_1\thicksim A_9}$ are related to the points {$A_1\thicksim A_9$ in Fig. \ref{extremerhds}.}}
\label{sample}
\end{center}
\end{figure}

To see the effect of the boundary lines, we draw a vertical line crossing all the boundary lines in Fig. \ref{extremerhds} and we let $(a^2,r_\text{h})$ moving along this line to see the change of the horizon situation of the black hole. On this vertical line, we choose one point in each different region (denoted by $A_1\thicksim A_9$), and for each point, we draw the horizon situation of the black hole in Fig. \ref{sample}.

{ Comparing Figs. \ref{Wregionds},  \ref{extremerhds} and \ref{sample},} we can find that when $(a^2,r_\text{h})$ is in region 1, region 2 or on line 1 in Fig. \ref{extremerhds}, the particle with critical angular momentum can reach the outer horizon of the Kerr-dS black hole. When $(a^2,r_\text{h})$ is on line 2, $r_\text{h}$ is the outer extremal horizon and $\frac{d R(r)}{dr}|_{r=r_\text{h}}=0$. By computing $\frac{d^2 R(r)}{dr^2}|_{r=r_\text{h}}$, it can be proved that $\frac{d^2 R(r)}{dr^2}|_{r=r_\text{h}}$ is positive. So When $(a^2,r_\text{h})$ is on line 2, the particle with critical angular momentum can reach the outer horizon of the Kerr-dS black hole.

As a summary, we find as long as $r_\text{h}$ is the outer horizon
of the Kerr-dS black hole, the particle with critical
angular momentum can always reach the horizon $r_\text{h}$.
This means the particle with critical angular momentum can
always reach the outer horizon of the Kerr-dS black hole
without constraints coming from the geodesic motion of the particle. { This is very different from the case of Kerr and Kerr-AdS black holes.}

{\subsection{From the Kerr-AdS case to the Kerr-dS case}}

{Here we analyze in detail that how the Kerr-AdS (Kerr-dS) case changes into the Kerr case when the cosmological constant changes from negative to positive.

When the cosmological constant turns from the negative to zero, Eq. (\ref{rhads}) becomes
\begin{equation}
r_{\text{h}}=a. \label{rhkerr}
\end{equation}
We denote $W_{\text{Kerr}}\equiv W_{\text{AdS}}|_{\Lambda\rightarrow 0}$. We draw $r_{\text{h}}=a$ in Fig. \ref{WregionKerr}. So we can see $r_{\text{h}}=a$ can serve as a boundary line. { Thus, the particle with critical angular momentum can reach Kerr black hole's outer horizon only if the Kerr black hole is extremal.}

 \begin{figure}

 \includegraphics[width=120mm]{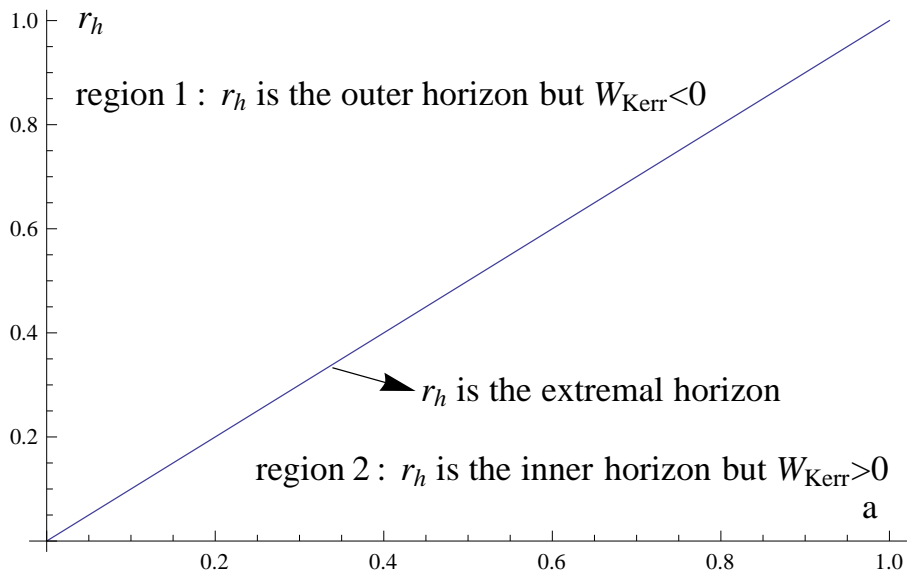}
 \caption{The shape of $r_h(a)$ for the case of the extremal Kerr black hole. This figure is similar to Fig. \ref{Wregionads}.}
 \label{WregionKerr}
 \end{figure}

When the cosmological constant turns from positive to zero, the right part of Eq. (\ref{rhds1}) also changes into $a$ and the right part of Eq. (\ref{rhds2}) becomes positive infinite.

Recalling Figs. \ref{Wregionads}, \ref{Wregionds} and  \ref{WregionKerr}, we find that the curved boundary line in Fig. \ref{Wregionads} will become the straight boundary line in Fig. \ref{WregionKerr} when the cosmological constant turns from negative to zero. And the straight boundary line in Fig. \ref{WregionKerr} will bend back in Fig. \ref{Wregionds}. So it can surround the region 1 in Fig. \ref{Wregionds} completely when the cosmological constant turns from zero to the positive. In this way, the region 1 in Fig. \ref{Wregionds} where $W_{\text{dS}}<0$ can be bounded and outside the boundary, $W_{\text{dS}}>0$ and $r_{\text{h}}$ can still be the outer horizon. This is why particles with critical angular momentum can reach the outer horizon of the Kerr-dS black hole without requiring the black hole to be extremal.

In fact, $\frac{d R(r)}{dr}|_{r=r_\text{h}}$ can be rewritten as
\begin{equation}
\frac{d R}{d r}|_{r=r_\text{h}} = -\frac{\left(a^2+E^2 r_\text{h}^2\right)}{a^2 r_\text{h}^2} \frac{d \Delta_r}{d r}|_{r=r_\text{h}}, \label{1-order}
\end{equation}
and when $\frac{d R}{d r}|_{r=r_\text{h}} = 0$,
\begin{equation}
\frac{d^2 R}{d r^2}|_{r=r_\text{h}} = \left(\frac{8 }{r^2}-\frac{1}{a^2}\frac{d^2\Delta _r|_{r=r_\text{h}}}{dr^2}|_{r=r_\text{h}}\right)E^2-\frac{1}{r^2}\frac{d^2\Delta _r}{dr^2}|_{r=r_\text{h}}. \label{2-order}
\end{equation}
For the Kerr-AdS or the Kerr black hole, on the outer non-extremal horizon, $\frac{d \Delta_r}{d r}>0$; and on the extremal horizon, $\frac{d \Delta_r}{d r}=0$ and $\frac{d^2\Delta _r}{dr^2}>0$. So as $-\frac{\left(a^2+E^2 r_\text{h}^2\right)}{a^2 r_\text{h}^2}<0$, $\frac{d R}{d r}$ cannot be positive on the outer non-extremal horizon. Thus the outer horizon has to be extremal, and as $\frac{d^2R}{dr^2}$ must be positive on the extremal horizon, from Eq. (\ref{2-order}), we can see that the parameters of the black hole and the particle must be confined. In fact, we can still get (\ref{addionads1}) and (\ref{addionads2}) in this way.

For the Kerr-dS black hole, on the outer non-extremal horizon, $\frac{d \Delta_r}{d r}<0$; and on the outer extremal horizon, $\frac{d \Delta_r}{d r}=0$ and $\frac{d^2\Delta _r}{dr^2}<0$. So from Eqs. (\ref{1-order}) and (\ref{2-order}), we can see that on the outer non-extremal horizon, $\frac{d R}{d r}$ is positive; and on the outer extremal horizon, $\frac{d R}{d r}=0$ and $\frac{d^2R}{dr^2}>0$. { Thus, the Kerr-dS black hole needs not to be extremal and there is no additional condition needed.}

From above analysis, { we know why the Kerr-AdS and Kerr cases are similar and why the Kerr-dS case is so different.}}

\section{Conclusion}
\label{SecConclusion}

In this work, we have analyzed the possibility that  Kerr-(anti-) de Sitter black holes could act as particle accelerators.
{ We find that the result is different from the case of Kerr black holes because of the non-vanishing cosmological constant in the background spacetime. In order for two particles to collide on
outer horizon of the Kerr, Kerr-AdS, or Kerr-dS black holes
and to reach arbitrary high CM energy, one and only one of the
colliding particles should have a critical angular momentum.
Besides, for the case of the Kerr black hole, it has to be extremal. For the Kerr-AdS one,
it has to be extremal,
and an additional condition should be satisfied.
However, for the case of the Kerr-dS black hole,
it does not need to be extremal and no additional condition
need to be satisfied. Hence, non-extremal Kerr-de Sitter black holes could also act
as particle accelerators with arbitrarily high CM energy, which is very different from the cases
of the Kerr and Kerr-AdS black holes.}
By analyzing how the Kerr-AdS (Kerr-dS) case changes into the Kerr case when the cosmological constant vanishes, we have seen exactly why the Kerr-dS case is so different.

\section*{Acknowledgements}

Y. Li is grateful to Dr. Pujian Mao for the valuable discussion. This work was supported by the National Natural Science Foundation of China (No. 11075065), the Doctoral Program Foundation of Institutions of Higher Education of China (No. 20070730055), the Fundamental Research Funds for the Central Universities (lzujbky-2010-171) and the Fundamental Research Fund for Physics and Mathematic of Lanzhou University (LZULL200907).

\end{document}